\documentclass[9pt,twocolumn,twoside]{osajnl}

\usepackage{hyperref}
\usepackage{siunitx}
\DeclareSIUnit{\belmilliwatt}{Bm}
\DeclareSIUnit{\dBm}{\deci\belmilliwatt}
\usepackage{physics}
\usepackage{braket}
\usepackage{epstopdf}
\usepackage{multirow}
\usepackage{dcolumn}

\DeclareSIUnit{\sqrthz}{\ensuremath{\sqrt{\text{\hertz}}}}

\journal{ol} 

\setboolean{shortarticle}{true}

\title{Heterodyne fiber interferometer for frequency-noise reduction and rapid wide-band tunability of a conventional laser source
}

\author[1]{Ashby P. Hilton}
\author[1,*]{Philip S. Light}
\author[1,2]{Lauris J. B. Talbot}
\author[1]{Andre N. Luiten}

\affil[1]{Institute for Photonics and Advanced Sensing (IPAS) \emph{and} School of Physical Sciences, University of Adelaide,
Adelaide, SA 5005, Australia}
\affil[2]{Centre d'Optique, Photonique et Laser, Universit\'{e} Laval, Qu\'{e}bec, QC, G1V 0A6, Canada}

\affil[*]{Corresponding author: philip.light@adelaide.edu.au}

\begin{abstract}
	Self-heterodyne fiber interferometers have been shown to be capable of stabilizing lasers to ultra-narrow linewidths and present an excellent alternative to high finesse cavities for frequency stabilization.
	In addition to suppressing frequency noise, these devices are highly tunable, and can be manipulated to produce high speed frequency sweeps over the entire range of the laser.
	We present an analytic approach for choosing a delay-line length for both optimal noise suppression and highest in-loop frequency sweep rate.
	Using this model we have implemented a fiber-based active Michelson interferometer as a frequency discriminator for a conventional diode laser and demonstrated a linewidth of \SI{700}{\hertz} over millisecond timescales.
	We also demonstrate a frequency scan rate of \SI{1}{\tera\hertz\per\second} and independently measure the maximum deviation in frequency from the linear sweep to be \SI{100}{\kilo\hertz}, predominantly limited by acoustic resonances in the apparatus.
\end{abstract}

\setboolean{displaycopyright}{true}

\begin{document}
	
	\maketitle

	\section{Introduction}
		\label{sect:Introduction}
		Fiber interferometers are capable of providing sensitive frequency discrimination for the reduction of laser linewidths.
		Long path length imbalances can be produced using a spool of optical fiber and these systems typically perform very well due to their high sensitivity and low inherent frequency noise floors \cite{Chen1989,Kefelian2009,Kong2015}.
		Key advantages are the ease of implementation of such a system and the locking flexibility, particularly in contrast to other conventional optical frequency references such as ultra-stable cavities and narrow atomic transitions.
				
		A primary limitation of conventional approaches to frequency stabilization, such as locks to an atomic transition or to an optical Fabry Perot (FP) resonator, is their inflexibility.
		These techniques only offer lock points at specific transitions, or at a number of discrete points.
		Without additional lasers and multiple frequency locks it is difficult to tune a stabilized laser beyond the range of an acousto-optic modulator (AOM) while retaining a narrow linewidth.
		In contrast, a self-heterodyne fiber interferometer (SHFI) allows locking of a laser at any optical frequency, as well as unbounded continuous tuning of this lock point over the full range of the laser at high speed \cite{Jiang2010,Crozatier2006,Lee2011}.
		
		Narrow laser sources with high scan speeds are of great importance for use in sensing protocols such as frequency modulated continuous wave (FMCW) reflectometry, for which these properties provide direct improvements in range and spatial resolution \cite{Geng2005}.
		The laser chirp produced by this technique is generated by the introduction of a frequency offset in the demodulation signal, and as such does not require manipulation of physical element of the apparatus.
		As the entire sweep is performed in-loop this protocol is highly robust and does not sacrifice the quality of the stabilization.
			
		Demonstrations of a similar technique using passive homodyne interferometers have produced sweep rates of up to \SI{100}{\tera\hertz\per\second} using carefully pre-calibrated kicks to the laser's frequency actuator \cite{Geng2005,Gorju2007,Satyan2009,Qin2019}.
		Although this can produce extremely high sweep rates, the disadvantage to this technique is that the frequency modulation must be predetermined in advance.
		The approach we use is able to maintain entirely in-loop performance and can produce arbitrary frequency ramps in real time without active feed-forward.
		
		A common property of laser stabilization techniques is the trade off between sensitivity and bandwidth, which for a SHFI are both set by the optical path imbalance.
		This choice determines the performance of the frequency stabilization, and maximum in-loop sweep rate.
		We present an approach to optimizing the fiber delay-line length for a given master laser frequency noise spectrum.
		Using an interferometer with parameters determined by this model we narrow the linewidth of a conventional ECDL to less than \SI{700}{\Hz}.
		We are also able to perform a frequency sweep over the full \SI{100}{\giga\hertz} mode-hop free range of the laser at a rate of \SI{1}{\tera\hertz\per\second}.
		
	\section{Principle of Operation}
		\label{sec:PrincipleOfOperation}
\begin{figure}
			\centering
			\fbox{\includegraphics[width=\linewidth]{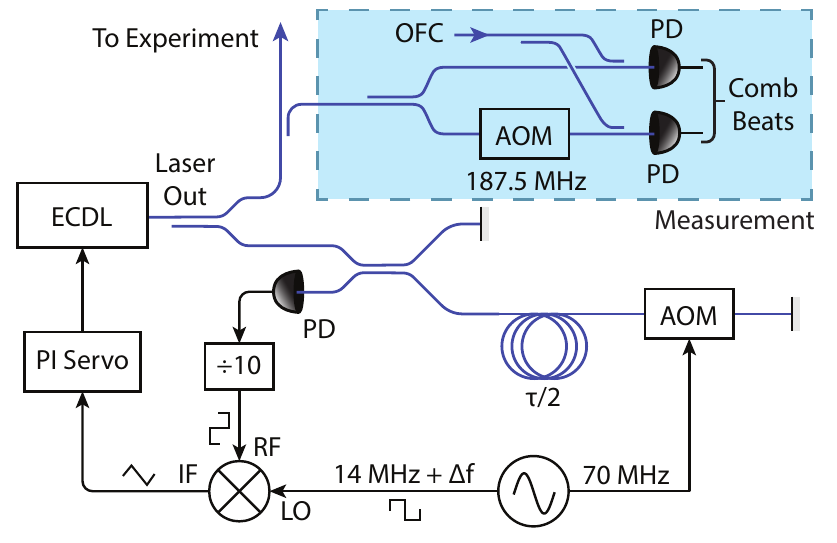}}
			\caption{
				\label{fig:Setup}
				Schematic of laser stabilization optics and electronics, showing both the Michelson interferometer and the optical frequency comb comparisons.
			}
		\end{figure}

		A self-heterodyne interferometer splits light into two arms: a short reference arm, and a long arms with relatively long round trip time delay $\tau$ which includes an element (typically an AOM) that produces a frequency shift.
		In the standard Michelson configuration each arm is retro-reflected, doubling the frequency shift, as depicted in \autoref{fig:Setup}.
		Recombination of the two electric fields generates a time varying intensity modulation that can be detected using a photo-diode.
		A change in laser frequency shifts the temporal phase of the intensity modulation at the output port of the interferometer.
		
		After amplification and filtering the signal is electronically divided down in frequency by a factor $N_\mathrm{div}$ for convenience and down mixed with a local oscillator at frequency $f_\mathrm{RF}=2\times f_\mathrm{AOM}/N_\mathrm{div}+\Delta f$, that is phase coherent with that supplied to the AOM.
		The resulting signal is given by
		\begin{equation}
			\label{eq:MixerOutput}
			V_\mathrm{err}\propto\sin\left(2\pi \nu\left(t\right) \tau/N_\mathrm{div} + 2\pi \Delta f t +\phi_0\right)
		\end{equation}
		where $\nu\left(t\right)$ is the optical frequency of the laser, and $\phi_0$ is the user controllable phase offset of the demodulation signal.
		This error signal has a periodicity in frequency akin to the free spectral range of a cavity given by $N_\mathrm{div}/\tau$, and when $\Delta f = 0$, produces a set of stationary zero-crossings suitable for use as lock points.
		Adjusting $\phi_0$ allows the user to shift the location of the zero crossings in frequency, while a linear phase ramp produced by introducing an offset in the demodulation frequency of $\Delta f$ causes these lock points to continuous shift in frequency at a rate given by
		\begin{equation}
			\label{eq:SweepRate}
			\frac{\dd{\nu\left(t\right)}}{\dd{t}} = -\frac{\Delta f \times N_\mathrm{div}}{\tau}
		\end{equation}
		until either $\Delta f$ is altered again, or the laser frequency actuator runs out of range.		 

	\section{Performance Model}
		\label{sec:PerformanceModel}
		The interferometer's transfer function for changes in optical frequency is given by
		\begin{equation}
			\label{eq:InterferometerTF}
			H_\mathrm{int}\left(f\right)=\frac{1-\exp\left(-i\,2 \pi f \tau \right)}{2\pi f}
		\end{equation}
		where the gain is DC is equal to $\tau$.		
		Typical implementations use a very long fiber delay line to maximize the sensitivity to frequency perturbations.
		However, the detection of a change in laser frequency is effectively limited by the time taken for the change to propagate through both arms of the device which produces a null in the sensitivity at a frequency $f_\mathrm{null}=1/\tau$.
		As the time delay is increased, the frequency of the null due shifts closer to DC, limiting the useful feedback bandwidth to the laser \cite{Chen1989}.
		While it is possible to design loop amplifiers that overcome the nulls in the interferometer sensitivity to achieve high bandwidth, these techniques require typically require complex high speed digital controllers \cite{Sheard2006,Gray2014}.
		Instead, we determine the optimum choice in $\tau$ for a given laser by analyzing its dependence of the performance of the stabilization servo.
	
		We choose a simple loop amplifier design of proportional + integrator (PI) given by 
		\begin{equation}
			\label{eq:PITF}
			H_\mathrm{PI}\left(f\right)=P \left(1+\frac{1}{i\, 4 \tau f}\right)
		\end{equation}
		with integrator cut-off frequency limited by half the \SI{-90}{\degree} interferometer bandwidth, $f_\mathrm{PI}=f_\mathrm{null}/4$.
		While it is possible to use additional integrator stages or even differential gain stages, for this demonstration we assume that a simple electronic system is desired.
		The choice of proportional gain is made such that the loop gain over all components result in under unity gain at the integrator corner to prevent loop oscillation.
		
		As one decreases the path length imbalance the corresponding increase in possible feedback bandwidth allows one to increase the integrator gain by the same factor.
		This counteracts the reduced sensitivity from the interferometer itself, and can continue until the loop bandwidth becomes limited by some other experimental detail, such as the speed of the laser frequency actuator itself.
		The true disadvantage of reducing the path length imbalanced is that the increased loop amplifier gain results in a stronger conversion of electronic input noise into laser frequency noise, limiting the attainable noise floor.
		Unlike many other locking techniques, photon shot noise on the detector is not typically a limiting noise source due to the relatively high optical power involved, as there is no problem with power build up as in a Fabry-Perot cavity, or saturation of an atomic transition.
		As such, the broadband limit to the lock performance is given by
		\begin{equation}
			\label{eq:NoiseFloor}
			S_{NF}\left(f\right)=\left(\frac{N_\mathrm{div}}{2 \tau V_\mathrm{pp}}\right)^2 \times S_{V,\,\mathrm{input}}\left(f\right)
		\end{equation}
		where $V_\mathrm{pp}$ is the peak to peak amplitude of the error signal, and $S_{V,\,\mathrm{input}}\left(f\right)$ is the combined voltage noise of the locking electronics before the loop amplifier.
				
		The predominant noise sources in a diode laser are frequency drift due to environmental sensitivity of the diode, low frequency flicker frequency noise, and broadband white frequency noise. 
		If one excludes slow drift, the frequency noise spectrum can be written as the polynomial law
		\begin{equation}
		\label{eq:UnlockedLaserNoise}
		S_{\Delta\nu\,\mathrm{unlocked}}\left(f\right)=\mathrm{d}_{-1}/f+\mathrm{d}_0
		\end{equation}
		where $\mathrm{d}_{-1}$ is the frequency flicker noise index in \SI{}{\hertz\cubed\per\hertz} and $\mathrm{d}_0$ is the white noise level in \SI{}{\hertz\squared\per\hertz}.
		By combining the gains of the servo amplifier, the unlocked laser noise, and the loop noise floor, we can calculate the expected performance of the stabilized laser as
		\begin{equation}
			\label{eq:LockedNoise}
			S_{\Delta\nu,\,\mathrm{locked}}\left(f\right) = S_{NF}\left(f\right) + \frac{S_{\Delta\nu,\,\mathrm{unlocked}}\left(f\right)}{1+\left|H_\mathrm{PI}\left(f\right)\right|^2 \left|H_\mathrm{int}\left(f\right)\right|^2}.
		\end{equation}.
		
		\begin{figure}
			\centering
			\fbox{\includegraphics[width=0.9\linewidth]{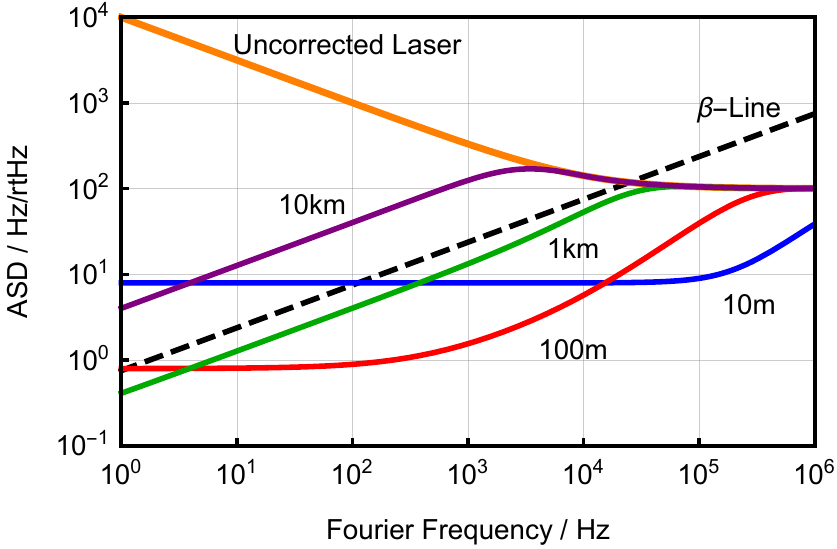}}
			\caption{
				\label{fig:Model}
				Model of the frequency noise of a diode laser when free-running, and when stabilised to an interferometer with a variety of delay line lengths.
			}
		\end{figure}
		We build a model of a conventional diode laser using values of $\mathrm{d}_0= \SI{e4}{\hertz\squared\per\hertz}$, ${\mathrm{d}_{-1}= \SI{e8}{\hertz\cubed\per\hertz}}$ for the free-running laser performance.
		The division factor is chosen to be 10 to allow us to use a \SI{20}{\mega\hertz} bandwidth waveform generator for the demodulation, and we assume an error signal with power of \SI{0}{\deci\belmilliwatt} and a servo input noise floor of  $S_{V,\,\mathrm{input}}\left(f\right) = \SI{e-14}{\volt\squared\per\hertz}$.
		In \autoref{fig:Model} we show the uncorrected laser noise as well as the calculated locked noise for four different choices of fiber delay line length: \SI{10}{\kilo\meter} (\SI{100}{\micro\second}), \SI{1}{\kilo\meter} (\SI{10}{\micro\second}), \SI{100}{\meter} (\SI{1}{\micro\second}), and \SI{10}{\meter} (\SI{100}{\nano\second}).
		The beta-separation line 
		\begin{equation}
			\label{eq:BetaLine}
			S_\beta\left(f\right)=\frac{8\ln 2}{\pi^2} f
		\end{equation}
		is also shown as a measure of coherence, where only frequency noise above the separation line and within the frequency bandwidth of the relevant measurement contributes to the spectral width of the laser \cite{DiDomenico2010}.
		
		The insight we draw from this model is that although one can generate a very high sensitivity interferometer using many kilometers of fiber in the delay line, the poor bandwidth can limit the gain available (purple curve).
		Alternatively, a short length (small $\tau$) will allow high bandwidth, but the low sensitively can result in a lock that is input noise limited at a high level over nearly the full frequency range (blue curve).
		For the modeled laser spectrum, a length of between \SI{100}{\meter} (red curve) and \SI{1}{\kilo\meter} (green curve) allows for enough bandwidth to suppress the frequency noise fo the laser to below the $\beta$-line down to below \SI{1}{\hertz}, effectively narrowing the laser linewidth.
		In reality one may be limited by the speed of the frequency actuator, which for the laser modeled is around \SI{300}{\kilo\hertz}.
		
		In addition to optimizing the lock performance, a careful choice of the delay length and beat division factor can vastly improve the ability to sweep the laser.
		We estimate the frequency slew rate possible using the slew rate calculation in \autoref{eq:SweepRate} $f_\mathrm{PI}$ where $\Delta f$ is taken to be $f_\mathrm{PI}$, the maximum frequency at which the interferometer lock can effectively suppress an introduced input error.
		The result provides an upper bound on the sweep rate attainable, given by
		\begin{equation}
			\label{eq:MaxSweep}
			\frac{\dd{\nu\left(t\right)}}{\dd{t}}\big|_{\Delta f=f_\mathrm{PI}} = \frac{c^2 N_\mathrm{div}}{16 n^2 l^2}
		\end{equation}
		where we have used physical units to highlight the inverse scaling with length squared.
		We note again that this will only hold while there is additional bandwidth to be gained by reducing the delay line length.
		For the model we presented earlier, this calculation gives a maximum frequency slew rate of $\sim\SI{2.6}{\tera\hertz\per\second}$.
		
	\section{Laser Stabilization}
		\label{sec:LaserStabilization}
		Based on this insight, we have implemented a basic demonstration of an optimized interferometer lock for a New Focus Vortex ECDL centered at \SI{894}{\nano\meter} for use in the spectroscopy of cesium.
		A few hundred microwatts of the laser output are tapped off from the main laser output and coupled into 50:50 fiber splitter.
		Half of the power is directed into a short fiber retro-reflector, forming the reference arm of the Michelson interferometer.
		The other arm of the splitter is connected to a \SI{100}{\meter} length of single-mode fiber, which is wound around a \SI{15}{\centi\meter} granite cylinder, forming the delay-line.
		At the end of the delay-line the light is coupled back out into free-space and double-passed through an AOM driven at $f_\mathrm{AOM}=\SI{70}{\mega\hertz}$.
		The returned light recombines with the reference beam and is incident on a fast photo-diode, nominally producing a steady-state heterodyne beat at \SI{140}{\mega\hertz}.
		
		The beat-note is divided down in frequency by a factor of $N_\mathrm{div} =  10$ and converted to a square wave using a combination of Hittite low-noise dividers, and then down-mixed with a square-wave from an arbitrary waveform generator referenced to the AOM driver.
		Use of a square wave for the mixer LO results in the error signal given by \autoref{eq:MixerOutput}, where the sine wave shape has been replaced with a triangle wave.
		This signal is amplified with a home-made low-noise PI controller and fed back to correct the laser frequency via current modulation, holding the laser stable.
		In addition the fast feedback is further integrated and fed back to the piezo-actuator for slow compensation to improve the locking range of the laser.

		The use of square wave demodulation has several advantages over conventional sine-wave demodulation.
		Not only does it produce a frequency discriminator with constant error slope over the entire peak to peak fringe, but the fixed output amplitude from the low-noise dividers also makes the feedback loop gain insensitive to perturbations in optical power in the interferometer.
		The resulting lock is incredibly robust as the servo retains its full gain for large excursions of laser frequency, and the linearity of frequency-to-voltage slope allows for accurate measurement of the in-loop performance of the lock.
		A downside is that the shape of the error signal is sensitive to filtering of higher harmonics that compose the square wave signals, and as such one must be careful to avoid using filters, amplifiers, or mixers with low cut-off frequencies.
		
		\begin{figure}
			\centering
			\fbox{\includegraphics[width=\linewidth]{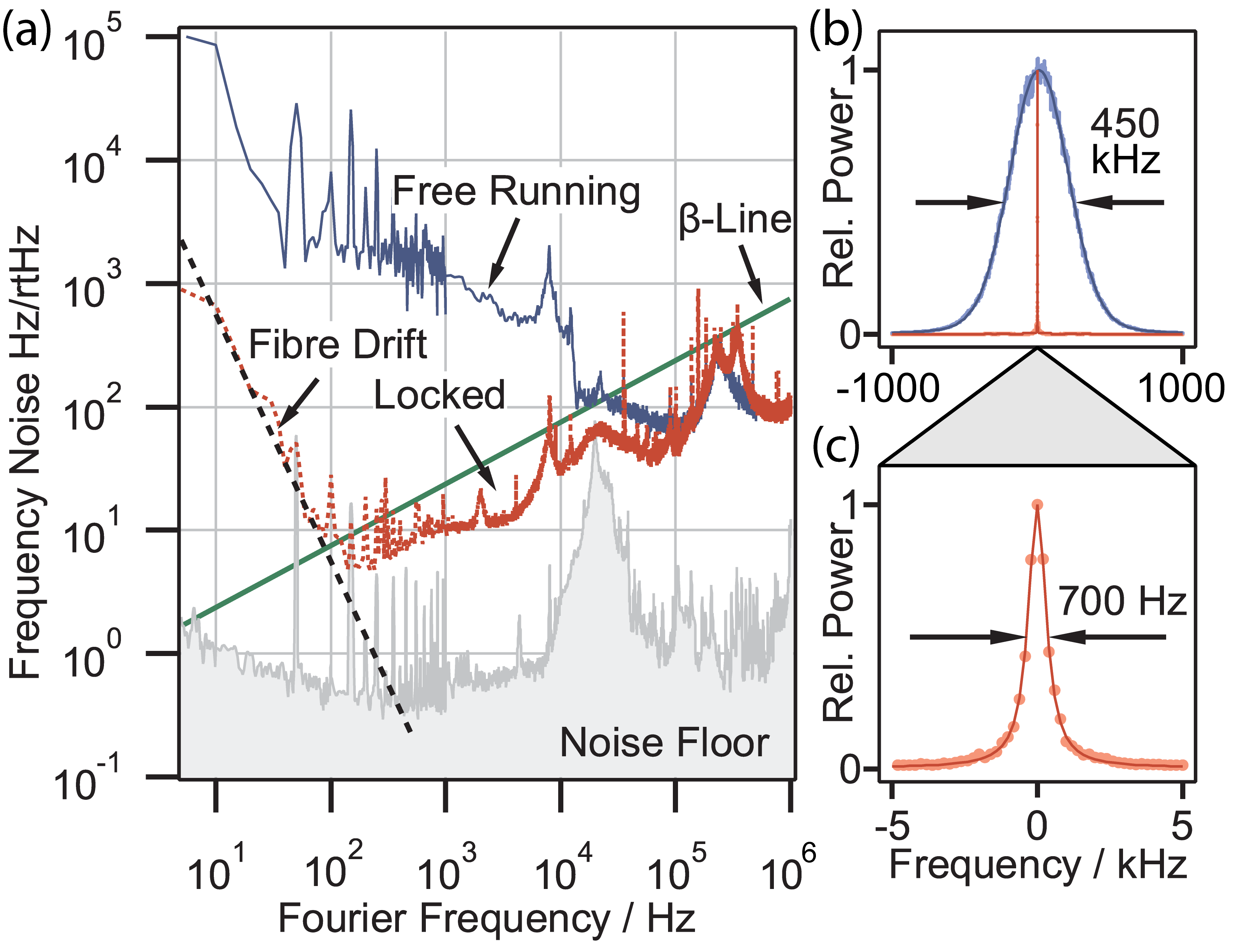}}
			\caption{
				\label{fig:Performance}
				(a) Frequency noise amplitude spectrum density of the laser, with the free-running laser (blue, solid data), the interferometer-stabilized laser (red, dotted data), and limiting electronic noise floor (gray, filled data).
				The green solid line is the $\beta$ separation line, and the black dashed line with slope $1/f^2$ is the raised low-frequency noise introduced by drift in the fiber delay-line length.
				(b) The laser power spectrum showing the full width of the unlocked laser, where the axis is expanded in (c) to show the spectrum of the laser while locked.
			}
		\end{figure}
		The performance of the lock is measured out of loop by beating the laser with an optical frequency comb.
		The comb has a repetition rate of $f_\mathrm{rep}=\SI{250}{\mega\hertz}$, and is itself phase locked to a fiber laser, which is in turn locked to a high finesse optical cavity.
		The beat note is measured using a fast photo-detector and filtered at half $f_\mathrm{rep}$ to isolate the beat with the closest comb mode.
		A high speed oscilloscope is used to collect a time trace of the beat which is processed digitally to produce an amplitude spectral density (ASD) of the laser frequency noise.
		
		Undergoing this process for both free-running and locked states of the laser, we demonstrate the performance of locking system as shown in \autoref{fig:Performance}.
		Part (a) demonstrates that this approach has suppressed frequency noise between \SI{100}{\hertz} and \SI{10}{\kilo\hertz} to below the $\beta$ separation line, narrowing the laser linewidth dramatically.
		The apparent reduced gain below \SI{100}{\hertz} is due to thermal drift of the interferometer, which is mapped onto laser frequency, and measured at $\approx\SI{16}{\kilo\hertz\per\second}$ during typical use.
		This is expected as the fiber spool is not temperature controlled, and is only mildly passively isolated.
		We expect the use of an active temperature servo would dramatically improve this source of drift \cite{Chen1989}, however we wished to demonstrate that complex vacuum pumped environments are not required to achieve relatively good performance.
		
		The free running laser spectrum is calculated over a \SI{5}{\milli\second} window and is well described by a Voigt profile with Gaussian full-width at half-maximum (FWHM) of \SI{450}{\kilo\hertz}, and Lorentzian component of \SI{60}{\kilo\hertz}, as shown in \autoref{fig:Performance} (b).
		When locked, the central feature is reduced to a \SI{700}{\hertz} Lorentzian FWHM with no discernible Gaussian component, shown in (c).

	\section{High Speed Linear Sweep}
		\label{sec:HighSpeedLinearSweep}
		 \begin{figure}
		 	\centering
		 	\fbox{\includegraphics[width=\linewidth]{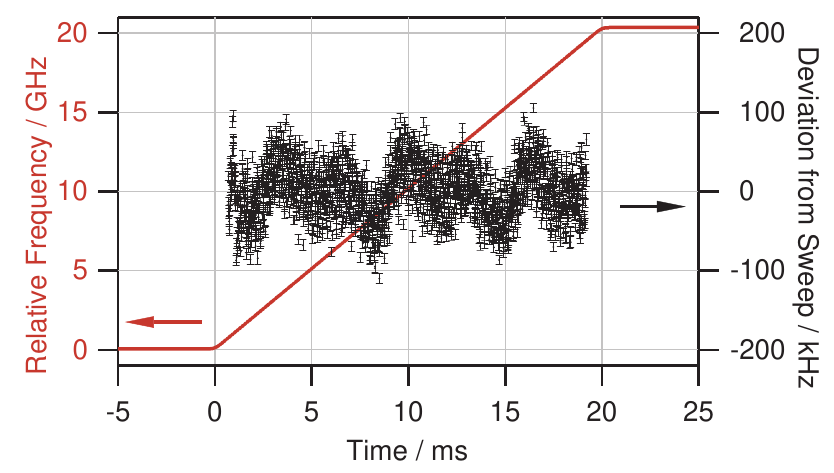}}
		 	\caption{
		 		\label{fig:ScanLinearity}
		 		Extracted relative frequency of the laser for a \SI{20}{\milli\second} duration sweep at \SI{1}{\tera\hertz\per\second}, with linear fit residuals on the right axis.
		 		The instantaneous frequency is binned in \SI{10}{\micro\second} segments. 
		 	}
		 \end{figure}
	 	As described in \autoref{sec:PrincipleOfOperation}, once locked we introduce a frequency offset to the local oscillator to produce a high speed linear sweep of the laser frequency.
	 	Using a \SI{100}{\kilo\hertz} offset we were able to achieve a sweep rate of \SI{1.0154891(2)}{\tera\hertz\per\second}.
	 	The laser stayed locked during the sweep over its full mode-hope-free range of \SI{100}{\giga\hertz},which we confirm using a wavemeter.
	 	
	 	We measure the laser frequency during the sweep using two comb beats as shown in the measurement section of \autoref{fig:Setup}, where for one beat the ECDL has been shifted by $3/4\times f_\mathrm{rep}$.
	 	Both beats are band-pass filtered between \SI{12.5}{\mega\hertz} and \SI{125}{\mega\hertz} to isolate the beat with a single comb mode.
	 	The frequency shift guarantees that when one comb beat approaches DC or $f_\mathrm{rep}/2$, the other beat is in the center of the filter window, allowing for good measurement over the entire comb spectrum.
	 	High speed oscilloscope traces are taken during the frequency sweep, and both instantaneous beat frequencies are calculated.
	 	As this process raises the noise at high fourier frequencies, we then filter and down-sample the data into \SI{10}{\micro\second} bins, to roughly match the order \SI{100}{\kilo\hertz} bandwidth of the system.
	 	Using both comb beats it is possible to unwrap the instantaneous frequency data and measure the performance of the sweep, shown in red on the left axis in \autoref{fig:ScanLinearity}.
	 	 	
	 	Unfortunately the spectral width of our optical frequency comb is currently limited to \SI{20}{\giga\hertz} centered on the cesium $D_1$ line, and as such we are only able to verify the linearity of the sweep over a fifth of the full range.
	 	The maximum deviation from the linear sweep over this range is found to be \SI{100}{\kilo\hertz}, while the root-mean-squared (RMS) deviation is \SI{35}{\kilo\hertz}.
		We note that this frequency deviation is independent of the relative frequency of the laser, and as such is independent on sweep length for a given frequency slew rate.
		As such we expect the maximum sweep error over the full mode-hop-free limited sweep range of the laser to be 1 part-per-million in the \SI{100}{\kilo\hertz} measurement bandwidth we have chosen.
		The time dependence of the linear fit residuals are shown as black markers on the right axis demonstrate clear periodic structure.
		Fourier analysis shows that the structure is entirely dominated by bright lines at \SI{160}{\hertz} and \SI{4}{\kilo\hertz} and their second harmonics.
		We attribute these bright lines to acoustic resonances in the interferometer delay-line that are mapped into real laser noise by the lock, which could be addressed through improved acoustic dampening.

	\section{Conclusion}
		\label{sec:Conclusion}
		Current demonstrations of fiber delay line interferometers typically use either extremely long delay arms to maximize sensitivity, or extremely short delays to maximize bandwidth.
		We present a guide to selecting a path imbalance length that is optimal for suppressing the frequency instability of the laser source.
		This guide extends to predict the maximum linear frequency chirp attainable in-loop using such a device.
		Using this model we have implemented a laser stabilization system that reduced the linewidth of a conventional ECDL to \SI{700}{\hertz} over a \SI{5}{\milli\second} window and have demonstrated an extremely high frequency scan rate of \SI{1}{\tera\hertz\per\second} over the entire mode-hop-free range of the laser using all in-loop control.
		Using a dual comb beat measurement system we calculate the sweep linearity to be \SI{200}{ppb}, sampled at \SI{10}{\micro\second} intervals, with maximum frequency deviation of \SI{100}{\kilo\hertz}.
		This slew rate and linearity is state of the art for entirely in-loop devices that do not require the application of pre-calibrated feed forward ramps.



\begin{thebibliography}{10}
		\newcommand{\enquote}[1]{``#1''}
		
		\bibitem{Chen1989}
		Y.~T. Chen, {\protect\JournalTitle{Applied Optics}} \textbf{28}, 2017 (1989).
		
		\bibitem{Kefelian2009}
		F.~K{\'{e}}f{\'{e}}lian, H.~Jiang, P.~Lemonde, and G.~Santarelli,
		{\protect\JournalTitle{Optics Letters}} \textbf{34}, 914 (2009).
		
		\bibitem{Kong2015}
		J.~Kong, V.~G. Lucivero, R.~Jim{\'{e}}nez-Mart{\'{i}}nez, and M.~W. Mitchell,
		{\protect\JournalTitle{Review of Scientific Instruments}} \textbf{86} (2015).
		
		\bibitem{Jiang2010}
		H.~Jiang, F.~K{\'{e}}f{\'{e}}lian, P.~Lemonde1, A.~Clairon1, and G.~Santarelli,
		{\protect\JournalTitle{Optics Express}} \textbf{18}, 3284 (2010).
		
		\bibitem{Crozatier2006}
		V.~Crozatier, G.~Gorju, F.~Bretenaker, J.-L. {Le Gou{\"{e}}t},
		I.~Lorger{\'{e}}, C.~Gagnol, and E.~Ducloux, {\protect\JournalTitle{Applied
				Physics Letters}} \textbf{89}, 261115 (2006).
		
		\bibitem{Lee2011}
		W.~K. Lee, C.~Y. Park, J.~Mun, and D.~H. Yu, {\protect\JournalTitle{Review of
				Scientific Instruments}} \textbf{82} (2011).
		
		\bibitem{Geng2005}
		J.~Geng, C.~Spiegelberg, and S.~Jiang, {\protect\JournalTitle{IEEE Photonics
				Technology Letters}} \textbf{17}, 1827 (2005).
		
		\bibitem{Gorju2007}
		G.~Gorju, A.~Jucha, A.~Jain, V.~Crozatier, I.~Lorger{\'{e}}, J.-L. {Le
			Gou{\"{e}}t}, F.~Bretenaker, and M.~Colice, {\protect\JournalTitle{Optics
				Letters}} \textbf{32}, 484 (2007).
		
		\bibitem{Satyan2009}
		N.~Satyan, A.~Vasilyev, G.~Rakuljic, V.~Leyva, and A.~Yariv,
		{\protect\JournalTitle{Optics express}} \textbf{17}, 15991 (2009).
		
		\bibitem{Qin2019}
		J.~Qin, L.~Zhang, W.~Xie, R.~Cheng, Z.~Liu, W.~Wei, and Y.~Dong,
		{\protect\JournalTitle{Optics Express}} \textbf{27}, 19359 (2019).
		
		\bibitem{Sheard2006}
		B.~S. Sheard, M.~B. Gray, and D.~E. McClelland, {\protect\JournalTitle{Applied
				Optics}} \textbf{45}, 8491 (2006).
		
		\bibitem{Gray2014}
		M.~B. Gray, T.~G. McRae, S.~Ngo, D.~A. Shaddock, and M.~T.~L. Hsu,
		{\protect\JournalTitle{23rd International Conference on Optical Fibre
				Sensors}} \textbf{9157}, 915726 (2014).
		
		\bibitem{DiDomenico2010}
		G.~{Di Domenico}, S.~Schilt, and P.~Thomann, {\protect\JournalTitle{Applied
				Optics}} \textbf{49}, 4801 (2010).
		
	\end{thebibliography}
\end{document}